\documentclass[preprint,pra]{revtex4}
\usepackage{amssymb,amsmath,graphicx,amscd,xcolor}
\usepackage{float}
\begin{document}

\title{Electric Field and Voltage Fluctuations in the Casimir Effect}

\author{L. H. Ford}
\email{ford@cosmos.phy.tufts.edu}
\affiliation{Institute of Cosmology, Department of Physics and Astronomy,
Tufts University, Medford, Massachusetts 02155, USA}

\begin{abstract}
The effects of reflecting boundaries on vacuum electric field fluctuations are treated. The presence of the boundaries can enhance these fluctuations
and possibly lead to observable effects. The electric field fluctuations lead to voltage fluctuations along the worldline of a charged particle moving
perpendicularly to a pair of reflecting plates, These voltage fluctuations in turn lead to fluctuations in the kinetic energy of the particle, which may
enhance the probability of quantum barrier penetration by the particle. A recent experiment by Moddel {\it et al} is discussed as a possible example
of this enhanced barrier penetration probability.
\end{abstract}

\maketitle
\baselineskip=14pt	

\section{Introduction}
\label{sec:intro}

The Casimir effect has become the topic of extensive theoretical and experimental work in recent years~\cite{KMM09}. 
In its original form, it is a force of attraction between a pair of perfectly reflecting plates due to modification of the
electromagnetic vacuum fluctuations. The presence of the plates modifies vacuum modes whose wavelengths are of the order 
of the plate separation, and shifts the energy of the vacuum state by an amount which is proportional to an inverse power
of the separation. The plates also modify local observables in the region between the plates. These observables can
include the energy density and the electric field correlation functions. The shifts in the electric field fluctuations can in
principle be detected by charged particles moving between the plates in the form of modified Brownian motion. This has
been the topic of several investigations in recent years~\cite{YF04,YC04,HL06,SW07,PF11,DR19}. For example, Ref.~\cite{YF04}
examined the shift in the mean squared components of the velocity of a charge moving parallel to a plate, and found that
this shift can be negative, which was interpreted as a small reduction in the quantum uncertainty.

In the present paper, this problem will be re-examined, with particular attention to particles moving perpendicular to one or two
plates. The motivation for this study is a recent experimental result by Moddel {\it et al}~\cite{Moddel1,Moddel2}, who found that the current
flowing through a metal-insulator-metal (MIM) interface can be very sensitive to the distance between the interface and an aluminum
mirror. These authors conjecture that this dependence may be related to the Casimir effect. The purpose of the present paper
is to explore this conjecture in more detail. The outline of the paper is as follows: The electric field correlation functions in the presence
of two parallel perfectly reflecting plates will be reviewed in Sec.~\ref{sec:corr}. This results will be used in Sec.~\ref{sec:volt-flucts}
to compute voltage fluctuations along a segment of a particle's worldline, which yield fluctuations in the particle's kinetic energy.
The   Moddel {\it et al}  experiment is reviewed in Sec~\ref{sec:experiment}, and the extent to which its results might explained 
by the modified electric field fluctuations is discussed. The paper is summarized in Sec~\ref{sec:final}. 

Lorentz-Heaviside units in which $\hbar = c =1$ are used throughout the paper, except as otherwise noted.

\section{Electric Field Correlation Functions}
\label{sec:corr}

We may define a vacuum correlation function for the Cartesian components of the quantized electric field operator ${\bf E}(t,{\bf x})$ as
$\langle E^i (t,{\bf x})\, E^j (t',{\bf x}') \rangle$. Here we are primarily interested in the shift in the correlation
functions due to the presence of mirrors, which is described by the renormalized function
\begin{equation}
\langle E^i (t,{\bf x})\, E^j (t',{\bf x}') \rangle_R = \langle E^i (t,{\bf x})\, E^j (t',{\bf x}') \rangle - \langle E^i (t,{\bf x})\, E^j (t',{\bf x}') \rangle_0\,, 
\end{equation}
where $\langle E^i (t,{\bf x})\, E^j (t',{\bf x}') \rangle$ is an expectation value in the Casimir vacuum with the mirrors present, and
$\langle E^i (t,{\bf x})\, E^j (t',{\bf x}') \rangle_0$ is an expectation value in the Minkowski vacuum without mirrors. We consider the 
Casimir geometry of two parallel, perfectly reflecting mirrors, one located at $z=0$ and the other at $z = a$. The correlation
functions for this geometry were calculated by Brown and Maclay~\cite{BM}, using an image sum method. We are especially
interested in the the case $i = j =z$, the correction function between the $z$-component of the electric field at two different spacetime
points which lie along a line perpendicular to the mirrors, so $x = x'$ and $y = y'$. In this case, the results of Ref.~\cite{BM} may be
used to show that
\begin{eqnarray}
\langle E^z(t,z)\, E^z (t',z') \rangle_R &=& \frac{1}{\pi^2 [(t-t')^2 - (z+z')^2]^2} \nonumber \\ 
&+& \frac{1}{\pi^2} {\sum_{n = -\infty}^{\infty}}' \biggl\{ \frac{1}{ [(t-t')^2 - (z-z' - 2 a n)^2]^2}  \nonumber \\ 
 &+&  \frac{1}{ [(t-t')^2 - (z + z' - 2 a n)^2]^2} \biggr\}  \, ,
 \label{eq:EzEz}
\end{eqnarray}
where the prime on the summation denotes that the $n = 0$ term is omitted.
In the limit that $a \rightarrow \infty$, we obtain the result for a single mirror
\begin{equation}
\langle E^z(t,z) E^z(t',z') \rangle_R = \frac{1}{\pi^2 [(t-t')^2 -(z+z')^2]^2}\,. 
\label{eq:one-plate-corr}
\end{equation}
 
Note that in all cases $\langle E^z(t,z) E^z(t',z') \rangle_R > 0$, meaning that the presence of the plates enhances the electric field fluctuations 
compared to those in empty space. Some physical effects of this enhancement will be the primary topic of this paper.

\section{Voltage and Particle Energy Fluctuations}
\label{sec:volt-flucts}

Consider a particle with electric charge $q$ moving in the $z$-direction, normal to the plates. The work done by the electric field when the particle
moves from $z=z_0$ to $z=z_0 + b$ along a spacetime path described by $z=z(t)$, or equivalently  $t = t(z)$ is
\begin{equation}
\Delta U = q \int_{z_0}^{z_0 +b} E^z(t(z),z)\, dz \,.
\end{equation}
The corresponding voltage difference is $\Delta V = \Delta U/q$. 
In the vacuum state, $\langle E^z \rangle = 0$, so the mean work vanishes, $\langle  \Delta U \rangle = 0$. However, the variance is nonzero and 
the contribution to the variance due to the presence of the plates may be written as 
\begin{equation}
\langle (\Delta U)^2  \rangle = q^2 \int_{z_0}^{z_0+b} dz \, \int_{z_0}^{z_0+b} dz'\, \langle E^z(t,z) E^z(t',z') \rangle_R 
\label{eq:variance}
\end{equation}
Assume that the particle moves at an approximately constant speed $v$, so its worldline may be described by $t = z/v$, or $t' = z'/v$.

\subsection{Single Plate Case}
\label{sec:one-plate}

First consider the case of one plate, so Eq.~\eqref{eq:one-plate-corr} applies.
The variance of the particle's energy due to the plate may now be written as
\begin{equation}
\langle (\Delta U)^2  \rangle = \frac{q^2 \, v^4}{\pi^2} \; I(z_0,b,v) \, ,
\label{eq:variance1}
\end{equation}
where
\begin{equation}
I(z_0,b,v) = \int_{z_0}^{z_0 +b} dz \int_{z_0}^{z_0 +b} dz'\, \frac{1}{[(z-z')^2 -v^2\, (z+z')^2]^2} \,.
\label{eq:I-def}
\end{equation}
This integral contains a second-order pole at points where $(z-z')^2 = v^2\, (z+z')^2$. It may be defined by writing
 the integrand as a second derivative:
 \begin{equation}
\frac{1}{[(z-z')^2 -v^2\, (z+z')^2]^2}  = \frac{\partial}{\partial z} \, \frac{\partial}{\partial z'}\, F(z,z')\,. 
\end{equation}
An explicit form for $F(z,z')$ is     
\begin{eqnarray}
F(z,z') = \frac{1}{128\, v^3\, (z z')^2} \; \Bigl[   8vzz' &+&  (1-v^2)(z^2 -z'^2) \, \Bigl( \log\{[(1+v)z'+(v-1)z]^2/\ell^2) \}   \nonumber \\
 &-& \log\{[(1+v)z+(v-1)z']^2/\ell^2 \}  \Bigr) \Bigr]\, ,
 \label{eq:F}
\end{eqnarray}
where $\ell$ is an arbitrary constant with the dimensions of length. Note that $F(z,z')$ is independent of the actual value of
$\ell$; if we rescale $\ell \rightarrow \mu \,\ell$, then $\mu$ cancels. Now $I(z_0,b,v)$  may now be expressed as
\begin{equation}
I(z_0,b,v)  = F(z_0+b,z_0+b) - F(z_0+b,z_0) - F(z_0,z_0+b) + F(z_0,z_0)\, .
\label{eq:I-exact} 
\end{equation}

In the limit that $v \ll  1$, this result becomes
\begin{equation}
I(z_0,b,v) \sim \frac{z_0^2 + (z_0+b)^2}{ 8 \, z_0^2 \, (z_0+b)^2\, v^2} +  \frac{(2z_0+b)^2 (2z_0^2+2 b z_0 -b^2)}{24 b^2 z_0^2 (z_0+b)^2} +   O(v^2)\,.
\label{eq:I-v-small} 
\end{equation}
The leading term, proportional to $v^{-2}$, is independent of $b$ when $b \alt z_0$,
\begin{equation}
I(z_0,b,v) \approx  \frac{1}{ 4 \,z_0^2\, v^2} \,.
\label{eq:I-b-small} 
\end{equation}
Both the factor of $v^{-2}$ and the lack of dependence upon $b$ in the above result may be traced to the singular nature of the integrand in Eq.~\eqref{eq:I-def}. 
In the limit that $v \rightarrow 0$,
this integrand approaches $1/(z-z')^4$ and the integral diverges. This leads to the result that $I(z_0,b,v) \propto 1/v^2$ for small
$v$. If the integrand in Eq.~\eqref{eq:I-def} were bounded, we would expect to find  $I(z_0,b,v) \propto b^2$ for small $b$ rather
than Eq.~\eqref{eq:I-b-small}. However, there is a limit to how small $b$ may be for fixed $v$, as we need to have the $O(v^0)$ term in 
Eq.~\eqref{eq:I-v-small} to be small compared to the $O(1/v^2)$ term. In the case that $b \ll z_0$, the former term becomes $1/(3 b^2)$,
which is sufficiently small provided that
\begin{equation}
b \agt \frac{2}{\sqrt{3}} \, v \, z_0 \,.
\label{eq:b-lower bound}
\end{equation}
The lack of $b$ dependence in Eq.~(\ref{eq:I-b-small}) arises because the contribution of the second-order pole in Eq.~(\ref{eq:I-def}) is independent of the length
of the integration interval so long as  $ v \, z_0 \alt b \alt z_0$

Note that when $v \ll 1$ but $b \gg z_0$, Eq.~\eqref{eq:I-v-small}  yields
\begin{equation}
I(z_0,b,v) \approx  \frac{1}{ 8 \,z_0^2\, v^2} \,,
\label{eq:I-b-large} 
\end{equation}
one-half of its value for small $b$. Furthermore, the decrease in the energy variance as $b$ increases is monotonic. This decrease
can be attributed to anti-correlated electric field fluctuations. 

We may now combine Eqs.~\eqref{eq:variance} and \eqref{eq:I-b-small} to write
\begin{equation}
\langle (\Delta U)^2  \rangle \approx \frac{q^2 \, v^2}{4\, \pi^2 \, z_0^2}  \, ,
\label{eq:variance2}
\end{equation} 
when $v \ll 1$ and $b \ll z_0$. In this limit, the root-mean-square energy fluctuation is
\begin{equation}
\Delta U_{rms} =   \sqrt{\langle (\Delta U)^2  \rangle} \approx  \frac{q \, v}{2\, \pi \, z_0} \,.
\label{eq:rms-E}
\end{equation}

Note that this energy fluctuation corresponds to a voltage fluctuation of
\begin{equation}
\Delta V_{rms} = \frac{1}{q}\, \Delta U_{rms} 
\end{equation}
along the worldline of the charged particle. This fluctuation is proportional to the speed $v$.
Remarkably, it is independent of the distance travelled, $b$, so long as
\begin{equation}
\frac{2}{\sqrt{3}} \, v \, z_0 \alt b \ll z_0\,.
\end{equation}

\subsection{Two Plate Case}
\label{sec:two-plate}

Now we turn to the case of two parallel plates, where the shift in the electric field correlation function is given by
Eq.~\eqref{eq:EzEz}. Again we consider a particle moving at constant speed $v$ from $z_0$ to $z_0 + b$, and write 
Eq.~\eqref{eq:variance} as 
\begin{equation}
\langle (\Delta U)^2  \rangle = \langle (\Delta U)^2  \rangle_{\rm one\, plate} + 
 \frac{q^2 \, v^4}{\pi^2} \;  {\sum_{n = -\infty}^{\infty}}'  [I_{2A}(n)+ I_{2B}(n)]\,.
 \label{eq:2p-var}
\end{equation}
Here $ \langle (\Delta U)^2  \rangle_{\rm one\, plate}$ is the result for a single plate, calculated in the previous subsection,
and we let
\begin{equation}
I_{2A}(n) =  \int_{z_0}^{z_0 +b} dz \int_{z_0}^{z_0 +b} dz'\, \frac{1}{[(z-z')^2 -v^2\, (z+z' -2 a n)^2]^2} \,,
\label{eq:I2a}
\end{equation}
and
\begin{equation}
I_{2B}(n) =  \int_{z_0}^{z_0 +b} dz \int_{z_0}^{z_0 +b} dz'\, \frac{1}{[(z-z')^2 -v^2\, (z- z'-2 a n)^2]^2} \,.
\label{eq:I2b}
\end{equation}

We may evaluate $I_{2A}(n)$ by noting that 
\begin{equation}
\frac{1}{[(z-z')^2 -v^2\, (z+z' -2 a n)^2]^2} = \frac{\partial}{\partial z} \, \frac{\partial}{\partial z'}\, F(z-a n, z'-a n)\,,
\end{equation}
so
\begin{eqnarray}
I_{2A}(n) &=&   F(z_0+b - a n, z_0+b - a n) - F(z_0+b- a n,z_0- a n)   \nonumber  \\
&-& F(z_0- a n,z_0+b- a n) + F(z_0- a n,z_0- a n)\, .
\label{eq:2a-exact} 
\end{eqnarray}
Here $F$ may be taken to have the form given in Eq.~\eqref{eq:F}.
In the limit $v \ll 1$, the quantity $I_{2A}(n)$  is also proportional to $1/v^2$ and takes the form
\begin{equation}
I_{2A}(n) \sim \frac{2 z_0^2 -4 a n z_0 +2 b z_0 +2 a^2 n^2 -2 a b n +b^2}{8 v^2 (a n -z_0)^2 (a n +b -z_0)^2} \,.
\end{equation}
As $b \rightarrow 0$, this approaches a nonzero value
 \begin{equation}
I_{2A}(n) \rightarrow \frac{1}{4 v^2 (a n -z_0)^2} \,.
\end{equation}

We evaluate $I_{2B}(n)$ by first expressing its integrand as
 \begin{equation}
 \frac{1}{[(z-z')^2 -v^2\, (z- z'-2 a n)^2]^2}  = \frac{\partial}{\partial z} \, \frac{\partial}{\partial z'}\, G(z,z') \,,
\end{equation}
where
\begin{eqnarray}
G(z,z') = \frac{1}{64\, (n a v)^3} \; \Bigl[   8 n a v &+&  [(1-v^2)(z -z')  + 2 n a v^2] \, \Bigl( \log\{[(1+v)(z' - z)  + 2 n a v]^2/\ell^2) \}   \nonumber \\
 &-& \log\{[(1-v)(z-z') + 2 n a v]^2/\ell^2 \}  \Bigr) \Bigr]\, ,
 \label{eq:G}
\end{eqnarray}
This leads to
\begin{equation}
I_{2B}(n) = G(z_0+b,z_0+b) - G(z_0+b,z_0) - G(z_0,z_0+b) + G(z_0,z_0)\, ,
\label{eq:2b-exact} 
\end{equation}
which has a form independent of $b$,  but proportional to $1/v^2$, when $v \ll 1$,
\begin{equation}
I_{2B}(n) \sim \frac{1}{ 4\, a^2\, v^2\, n^2} \,. 
\end{equation}

For the case $v \ll 1$ and $b \ll z_0$, we may write Eq.~\eqref{eq:2p-var} as
\begin{equation}
\langle (\Delta U)^2  \rangle = \langle (\Delta U)^2  \rangle_{\rm one\, plate} + 
 \frac{q^2 \, v^2}{4 \pi^2 a^2} \;  {\sum_{n = -\infty}^{\infty}}'  \left[  \frac{1}{n^2} + \frac{1}{(n +z_0/a)^2}  \right]\,.
 \label{eq:2p-var2}
\end{equation}
The sums may be evaluated in closed form using 
\begin{equation}
 {\sum_{n = -\infty}^{\infty}}'  \frac{1}{n^2} = 2 \zeta(2) = \frac{\pi^2}{3}\,,
 \end{equation}
 where $\zeta$ is the Riemann zeta-function, and~\cite{GR}
\begin{equation}
\sum_{n = 1}^\infty \left[ \frac{1}{(n +x)^2} + \frac{1}{(n -x)^2} \right] = -\frac{1}{x^2} +\pi^2 \csc^2 (\pi x) \,.
\label{eq:sum}
\end{equation}
These identities lead to our result for the particle energy variance in the two-plate case
\begin{equation}
\langle (\Delta U)^2  \rangle = \frac{q^2 \, v^2}{12 a^2} \left[ 1 + 3 \csc^2\left(\frac{ \pi \, z_0}{a} \right) \right]\,.
 \label{eq:2p-var-final}
\end{equation}
Note that the contribution of the $1/x^2$ term in Eq,~\eqref{eq:sum} has cancelled the $\langle (\Delta U)^2  \rangle_{\rm one\, plate}$
term in   Eq,~\eqref{eq:2p-var2}. Furthermore, $\langle (\Delta U)^2  \rangle$ is symmetric about the midpoint, $z_0 = a/2$. In the limit that
$z_0 \ll a$, it reduces to the one-plate result, Eq.\eqref{eq:variance2}. In the limit that $a-z_0$ is small, we have
\begin{equation}
\langle (\Delta U)^2  \rangle \sim \frac{q^2 \, v^2}{4 \pi^2  (a-z_0)^2} \,,
 \label{eq:2p-var-far-plate} 
\end{equation}
which is the one-plate result due to the plate at $z=a$.  

Note that here and in the previous subsection, we have assumed sudden switching at $z = z_0$ and $z = z_0 + b$ in expressions such as 
Eqs.~\eqref{eq:I2a} and \eqref{eq:I2b}.  This still produces finite results, as we are interested in the difference between the Minkowski and
Casimir vacua, which is sensitive to modes whose wavelength is of the order of the distance to the plates. Had we been dealing with 
effects in the Minkowski vacuum alone, sudden switching can produce infinite results, as modes of arbitrarily short wavelength could
contribute. Here we are justified in using sudden switching if the timescale for the onset of the coupling of the modified vacuum fluctuations 
to the charge is short compared to other timescales. If this is not the case, then the switching needs to be modeled by a smooth function,
as discussed in Refs.~\cite{SW07,DR19}.

\subsection{Some Estimates}
\label{sec:estimates}

Here we wish to make some numerical estimates of the magnitude of the voltage or particle energy fluctuations. Note that in all cases
studied above, the energy variance is proportional to $q^2 v^2$ and inversely proportional to the square of the distance between the starting point
and the nearest plate. Thus we may use the one-plate result, Eq,~\eqref{eq:rms-E}, for the root-mean-square energy fluctuation as an illustration.
Let $K = \frac{1}{2} m v^2$ be the particle's kinetic energy. In the case where the particle is an electron, we may write
\begin{equation}
\Delta U_{rms} =     \frac{e}{ \pi \, z_0} \, \sqrt{\frac{K}{ 2 m} } \approx 1.9 \times 10^{-4} {\rm eV}\, \sqrt{\frac{K}{1 \rm eV}}\, \left(\frac{100 \rm nm}{z_0} \right)\,.
\label{eq:rms-E2}
\end{equation}
Thus the magnitude of the energy fluctuations increases as the square root of $K$, but the fractional fluctuation, $\Delta U_{rms}/K$, is inversely proportional to
$\sqrt{K}$. In  any case, the energy and hence velocity fluctuations are relatively small for non-relativistic particles. This justifies our assumption that $v$ remains
approximately constant.

\subsection{Electric Field Fluctuations and Quantum Tunneling}
\label{sec:tunnel}

It is well known that a quantum particle can tunnel through potential barrier even when its energy is below the maximum of the barrier. It is also well known
that nonzero temperature can  enhance the rate at which particles can pass over the barrier. This arises not from quantum tunneling, but rather from the 
fraction of particles in the tail of the Boltzmann distribution that have enough energy to fly over the barrier classically. It is less well known that even at zero
temperature, quantum electric field vacuum fluctuations can enhance the tunneling rate compared to that predicted in single particle quantum mechanics~\cite{FZ99,HF15}.
In quantum electrodynamics, this effect arise as a one-loop radiative correction to the tree-level scattering amplitude for an electron to scatter from a potential
barrier. The basic physical process may be understood as follows: when the electron is in the vicinity of the barrier, it is equally likely to receive forward and
backward kicks from the electric field vacuum fluctuations. However, the net effect of the forward kicks is greater, so the tunneling rate is increased by the field 
fluctuations. The root-mean-squared energy fluctuation of an electron with initial kinetic energy $K$ while passing a barrier of width $a$ is found in
Ref.~\cite{HF15} to be
\begin{equation}
\Delta U_{MV} \approx \frac{e^2 \,  K}{m^2 \, a^2}\,.
\end{equation}
Here the effect is due to vacuum electric field fluctuations in the Minkowski vacuum of empty space.
This may be compared to the effect in the Casimir vacuum given by Eq.~\eqref{eq:rms-E2} to find
\begin{equation}
\frac{\Delta U_{rms}}{\Delta U_{MV} }  = 
\frac{a^2 \, m^{3/2}}{\pi \, e\, z_0\, \sqrt{K}} \approx 1.9 \times 10^4 \, \left(\frac{a}{1 {\rm nm}} \right)^2 \, \left(\frac{100 \rm nm}{z_0} \right)\, \sqrt{\frac{1 \rm eV}{K}} \,.
\end{equation}
For a wide choice of parameters, $\Delta U_{rms}  \gg \Delta U_{MV}$, so the effect due to the presence of plates is much larger than the effect in empty space.
The actual increase in tunneling rate due to the  presence of plates will depend upon the relative magnitudes of $\Delta U_{rms}$ and $|K - V_{max}|$, where
$ V_{max}$ is the maximum value of the potential. When these two quantities become comparable, the increase can be very large.

\subsection{Effects of Finite Reflectivity}
\label{sec:Finite}

So far, we have assumed perfectly reflecting plates. For a metal, this is a good approximation for electromagnetic waves with angular frequencies below the
plasma frequency, $\omega_p$,  but not for higher frequency modes. Thus, in a regime where the dominant contribution to a Casimir effect comes from modes
where $\omega \alt   \omega_p$, we can expect the results assuming perfect reflectivity to be a reasonable approximation. This is illustrated in calculations of
the mean squared electric field, $\langle E^2(z) \rangle$ at a distance $z$ from a single plate~\cite{SF02}. In the limit that $\omega_p \, z \agt 1$, we have
\begin{equation}
\langle E^2(z) \rangle \sim \frac{3}{16 \pi^2 \, z^4}\,.
\label{eq:E2-large-z}
\end{equation}
This is also the result at all values of $z$ for a perfectly reflecting plate, as may be found from the results in Ref.~\cite{BM}. In the case that $\omega_p \, z \alt 1$,
the mean squared electric field becomes
\begin{equation}
\langle E^2(z) \rangle \sim \frac{\sqrt{2}\, \omega_p}{32 \pi \, z^3}\,.
\label{eq:E2-small-z}
\end{equation}
Note that finite reflectivity modifies the singular behavior of $\langle E^2(z) \rangle$ as $z \rightarrow 0$, but does not remove it. This implies that another physical
cutoff is required. One possibility is that the assumption of an exactly smooth plane is too strong, and that surface roughness provides this cutoff.  For our purposes,
we need not answer this question, but rather confine the use of the perfectly reflecting results to the region where $z > 1/\omega_p$. 

Note that results such as Eq.~\eqref{eq:E2-small-z} give local expectation values, not correlation functions. A more detailed study of the latter in the presence of
boundaries with finite reflectivity is needed. Until such a study has been performed, it is not clear how sensitive results such as Eq~\eqref{eq:rms-E} are to finite
reflectivity.

\section{A Cavity-Induced Current Experiment}
\label{sec:experiment}

\begin{figure}[htbp]
\includegraphics[scale=0.25]{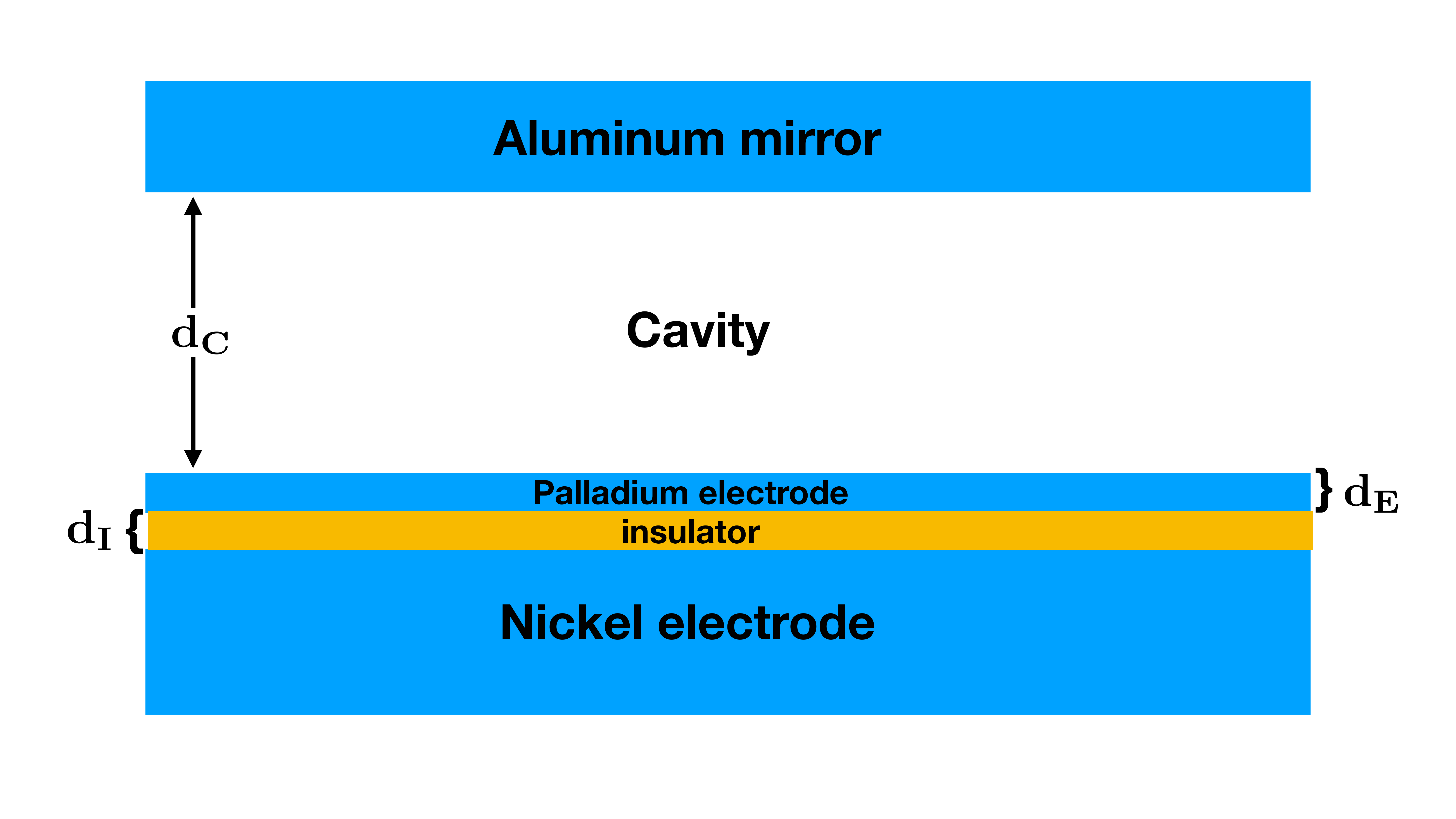}
\caption{An optical cavity of thickness $d_C$ is bounded by an aluminum mirror and a metal-insulator--metal (MIM) interface. The latter consists of a nickel electrode, a palladium
electrode of thickness $d_E$, and a layer of insulator of thickness $d_I$.}
\label{fig:cavity}
\end{figure}

Here we briefly summarize a recent experiment by Moddel {\it et al}~\cite{Moddel1,Moddel2}. This experiment involves electrical current in a metal-insulator-metal interface,
which is adjacent to a cavity, as illustrated in Fig.~\ref{fig:cavity}, which is adapted from Fig.~1 in Refs.~\cite{Moddel1,Moddel2}. A potential difference $V_0$ is imposed between
the palladium and nickel electrodes, which causes a current $I$ to flow through the layer of insulator separating the two electrodes. On the far side of the palladium electrode,
there is an optical cavity of thickness $d_C$, and beyond the cavity an aluminum mirror is located. The key result, illustrated in Fig.~3a of Ref.~\cite{Moddel1} or 
Fig.~4a of Ref.~\cite{Moddel2}, is that
the magnitude of $I$, for fixed $V_0$, is inversely related to the cavity thickness $d_C$. In effect, the electrical resistance of the insulator layer decreases as $d_C$ decreases.

The cavity is filled with a transparent dielectric, polymethyl methacylate (PMMA), and has various thicknesses, $d_C = 33 {\rm nm}, 79 {\rm nm}, 230 {\rm nm}, {\rm and}\; 1100 {\rm nm}$.
The insulator layer consists of $1.3  {\rm nm}$ aluminum oxide $Al_2 O_3$ and $1  {\rm nm}$  of nickel oxide, for a net thickness of $d_I = 2.3  {\rm nm}$. 
The palladium electrode has a thickness of $d_E = 8.3  {\rm nm}$, the aluminum mirror 
of $150  {\rm nm}$, and the nickel electrode  has a thickness of either $38  {\rm nm}$ or $50  {\rm nm}$. The plasma frequencies of aluminum, palladium, and nickel are, respectively,
$\omega_p(Al) = 15 {\rm eV}$,  $\omega_p(Pd) = 7.4 {\rm eV}$, and $\omega_p(Ni) =  9.5 {\rm eV}$. The corresponding length scales are $1/\omega_p(Al) = 14 {\rm nm}$,
$1/\omega_p(Pd) = 22 {\rm nm}$, and $1/\omega_p(Ni) = 27 {\rm nm}$,

The measured current flows through the layer of insulator, possibly by quantum tunneling. The distance of this layer from the aluminum mirror is $z_0 \approx d_C$, and in all cases
$\omega_p(Al) \, z_0 >1$. Hence we may approximate the aluminum mirror as a perfect mirror for the purpose of estimating its effect on the electric field fluctuations at the MIM location.
The palladium electrode may be viewed as approximately transparent because $\omega_p(Pd) \, d_E \ll 1$. The effect of the nickel electrode is difficult to assess. It is too close
to the insulator to be treated as a perfect mirror, as $\omega_p(Ni) \, d_I \ll 1$, but its effect could be significant.

The applied potential  differences between the nickel and palladium electrodes in Ref.~\cite{Moddel2} are of the order of $0.1 {\rm mV}$, which would produce kinetic energies of the order of 
$K \approx 10^{-4} \, {\rm eV}$ for freely accelerating electrons. We will use this value of $K$ and set $z_0 = 33 {\rm nm}$ in Eq.~\eqref{eq:rms-E2} to estimate the magnitude
of the electron energy fluctuations to be of order
\begin{equation}
\Delta U_{rms} \approx 0.06\, K \approx 6 \times 10^{-6}  {\rm eV} \,.
\end{equation}
The corresponding potential  differences in Ref.~\cite{Moddel1} are of the order of $0.2 {\rm V}$, which leads to the estimate
\begin{equation}
\Delta U_{rms} \approx 0.0013\, K \approx 2.5 \times 10^{-4}  {\rm eV} \,.
\end{equation}
It is unclear whether either of these are large enough to explain the results for the current flowing through the insulator discussed Refs.~\cite{Moddel1,Moddel2}. A more detailed
model of the potential barrier involved is needed. It should also be noted that that Ref.~\cite{Moddel2}  seems to find a small current even at zero applied voltage. There does not 
seem to be a plausible explanation of this effect in terms of vacuum electric field fluctuations.

\section{Summary}
\label{sec:final}

In this paper, we have examined the effects of vacuum electric field fluctuations on a charged particle moving perpendicularly to one or two perfectly reflecting plates.
This was done by integrating the electric field correlation function in the Casimir vacuum along a segment of the particle's worldline, and result in expressions describing
fluctuations in the voltage difference along the segment, and in the particle's energy. We then considered the possibility that these energy fluctuations could be linked to 
enhanced quantum tunneling through a potential barrier. The possibility that this effect has already been observed is discussed.

\begin{acknowledgments} 
I would like to thank G. Moddel for calling Refs.~\cite{Moddel1,Moddel2}  to my attention, and for helpful correspondence.
This work was supported in part  by the National Science Foundation under Grant PHY-1912545.

\end{acknowledgments}

\end{document}